\documentclass[prd,preprint,superscriptaddress,preprintnumbers,eqsecnum,showpacs,nofootinbib,nobibnotes]{revtex4-1}
\usepackage{amsfonts,amsmath,amssymb,bm}
\usepackage[table]{xcolor}
\usepackage{graphicx} 
\usepackage{mathtools}
\usepackage{boondox-cal}
\usepackage{slashed}
\usepackage{xcolor}

\usepackage[utf8]{inputenc}
\usepackage{slashed}

\newcommand{\be}{\begin{equation}}
\newcommand{\bea}{\begin{eqnarray}}
\newcommand{\ba}{\begin{align}}
\newcommand{\ee}{\end{equation}}
\newcommand{\eea}{\end{eqnarray}}
\newcommand{\ea}{\end{align}}

\definecolor{zero2}{rgb}{0.88,0.88,.88}

\def\1eq#1{Eq.~(\ref{#1})}

\def\2eqs#1#2{Eqs.~(\ref{#1}) and~(\ref{#2})}
\def\3eqs#1#2#3{Eqs.~(\ref{#1}),~(\ref{#2}) and~(\ref{#3})}
\def\4eqs#1#2#3#4{Eqs.~(\ref{#1}),~(\ref{#2}),~(\ref{#3}) and~(\ref{#4})}

\def\s#1{{\scriptscriptstyle #1}}

\def\G{\Gamma}

\def\s{\mathcal{s}}

\def\hphi0{{\hat\phi}_0}

\def\Mp2{{M'}^2}
\def\dEulS{{\cal D}^{M^2}_z}
\def\dEulV{{\cal D}^{\Mp2}_{z'}}
\def\lS{{\cal l}_h}
\def\lV{{\cal l}_a}

\makeatletter
\def\user@resume{resume}
\def\user@intermezzo{intermezzo}
\newcounter{previousequation}
\newcounter{lastsubequation}
\newcounter{savedparentequation}
\def\CT@@do@color{%
	\global\let\CT@do@color\relax
		\@tempdima\wd\z@
		\advance\@tempdima\@tempdimb
		\advance\@tempdima\@tempdimc
		\advance\@tempdimb\tabcolsep
		\advance\@tempdimc\tabcolsep
		\advance\@tempdima1.5\tabcolsep
	\kern-1.5\@tempdimb
	\leaders\vrule
	\hskip\@tempdima\@plus  1fill
	\kern-1.5\@tempdimc
	\hskip-\wd\z@ \@plus -1fill }
\makeatother

\begin{document}

\title{
Decoupling Limits in  Effective Field Theories\\
via Higher Dimensional Operators
}

\date{February 11, 2024}

\author{A. Quadri}
\email{andrea.quadri@mi.infn.it}
\affiliation{INFN, Sezione di Milano, via Celoria 16, I-20133 Milano, Italy}

\begin{abstract}
\noindent
The non-decoupling effects of heavy scalars and vector fields play an important role
in the indirect search for Beyond the Standard Model (BSM) physics at
the LHC.
By exploiting some new differential equations for the 1-PI
amplitudes,  we show that such non-decoupling effects are absent 
for quite a general class of effective field theories involving dimension six
two-derivative and dimension eight
four-derivative operators, once the resummation in  certain BSM couplings is taken into account and 
some particular regimes of the relevant couplings are considered.
\end{abstract}


\maketitle

\section{Introduction}

In the absence of any direct signal of new physics at the LHC, the~search for
Beyond the Standard Model (BSM) effects can be addressed within an Effective Field Theory
(EFT) approach~\cite{Manohar:2018aog,Efrati:2015eaa,Biekotter:2014gup,Crivellin:2014qxa,Weinberg:1980wa,Cepeda:2019klc,Buchmuller:1985jz,Grzadkowski:2010es,Brivio:2017vri,LHCHiggsCrossSectionWorkingGroup:2013rie,LHCHiggsCrossSectionWorkingGroup:2016ypw},
either in terms of Standard Model Effective Theories (SMEFT) or 
 Higgs Effective Field Theory (HEFT) (see e.g.,~the reviews in references
~\cite{Brivio:2017vri,Dobado:2019fxe}).

The high-energy dynamics can, in fact, induce measurable effects in low-energy observables,
despite the fact that new particles are too heavy to be directly detected in~experiments.

This is due to the so called non-decoupling effects (see e.g.,~\cite{Herrero:1994iu,Bilenky:1993bt,Dittmaier:1995cr,Dittmaier:2021fls,Arco:2023sac,Bhattacharyya:2014oka}), induced by loop corrections,
that survive in the large mass limit of the BSM particles living in the theory 
 assumed to be valid at high~energies.

There is no fundamental reason why the latter theory should be power-counting renormalizable.
Therefore, it is interesting to investigate whether some higher-dimensional operators 
in that theory might affect the non-decouping contributions to the low-energy physical observables one can directly measure at~colliders.

In the usual EFT treatment, computations are mostly limited to the first few terms in the small
coupling expansion.
In the present study, we will, on the contrary, consider a particular set of dimension six
two-derivative and dimension eight four-derivative operators, for~which a full resummation is~possible. 

The impact of these operators is quite dramatic, since, in some particular regimes, a
 complete decoupling of high-energy dynamics from the low-energy observables~occurs.

This result holds true for the fully resummed amplitudes that exhibit a qualitatively
different behavior from their small coupling~expansion.

The technical tool that allows one to study such a regime is the use of dynamical (i.e.,
propagating inside loops) gauge-invariant~variables.

The construction of the gauge-invariant dynamical counter-part of a 
scalar particle has been studied in~\cite{Quadri:2016wwl,Binosi:2022ycu,Binosi:2019nwz,Binosi:2020unh}. 

In the present study, we extend the analysis
to the case of vector fields. For~the sake of simplicity, we will only work
in the Landau gauge. The~ formalism in a generic $R_\xi$-gauge is technically more 
involved but does not change the physical content of the analysis. This will be presented~elsewhere.

The treatment of fermions will also be presented in a separate publication,
the reason being that, for fermions, it is not always possible to carry out a 
field transformation implementing a change in variables to their gauge-invariant
description. 
This can be achieved only
if the fermionic fields have some specific charge under the relevant gauge group, allowing for the building of a gauge-invariant combination out of the fermionic multiplet and the scalar containing the Higgs~mode. 

This is at variance with scalars and gauge fields, for~which the construction is always possible, provided that spontaneous symmetry breaking happens in the model at~hand.

We will work within the Algebraic Renormalization approach~\cite{Piguet:1995er,Grassi:1999tp,Grassi:2001zz}.

The key remark is that some
operators, that in the ordinary formalism modify both the quadratic and the interaction terms in the Lagrangian, are represented by purely quadratic
contributions if gauge-invariant variables are~used.

Therefore, they only affect propagators and, consequently, in~some particular cases,
one can write down a differential equation controlling the dependence of
the one-particle irreducible (1-PI) amplitudes on the BSM couplings~\cite{Binosi:2022ycu}.

These differential equations, in turn, can be exactly solved and lead to 
homogeneous Euler functions in the relevant couplings.
This result holds true for all orders in perturbation theory and provides
useful information on the structure of the fully resummed~amplitudes.

In particular, one can easily identify some regimes in which complete decoupling happens. In~those regimes, the small coupling expansion does not make any~sense.

From a physical point of view, these results cast a shadow on the feasibility of extracting physical information from non-decoupling effects, at~least within perturbation theory. It might, in fact, be that the physically relevant high-energy
dynamics are affected by the presence of such higher-dimensional~operators.

In that case, perturbative computations, that are limited to the first few terms in the small coupling expansion, are quite misleading, since they point to low-energy effects that are, indeed, not present in the full~theory.

This study is organized as follows. In~Section~\ref{sec.higgs.portal}, we set our notations and consider, for illustration purposes, a simple Higgs portal model, connecting a SU(2) spontaneously broken gauge theory at high energy to the SM.
In Section~\ref{sec.giv}, the gauge-invariant variables for the scalar and the vector fields are constructed.
In Section~\ref{sec.limit}, the differential equations for the 1-PI amplitudes
are derived and the non-decoupling limit is analyzed.
Finally, conclusions are presented in Section~\ref{sec.conclusions}.
Appendix~\ref{app.model} collects the functional identities of the theory, while
Appendix~\ref{app:landau} contains the derivation of the propagators in the Landau~gauge.

\section{A Simple Higgs Portal Model}\label{sec.higgs.portal}

{For the sake of definiteness, we will discuss a simple Higgs portal model with Lagrangian 

\begin{align}
    {\cal L} = {\cal L}_{SM} + {\cal L}_{ext} \, ,
\end{align}
where 
 ${\cal L}_{SM}$ is the
SM Lagrangian and ${\cal L}_{ext}$
is given by 
\begin{align}
{\cal L}_{ext} = & -\frac{1}{4} G_{a\mu\nu}^2 +
\frac{1}{4} 
{\rm Tr}
(D^\mu \phi)^\dagger D_\mu \phi -
\frac{\lambda}{2} \Big [ \frac{1}{2} {\rm Tr} (\phi^\dagger \phi) - v^2 \Big ]^2 \nonumber \\
& + g_1 {\rm Tr} \Big [ (\phi^\dagger \phi) - v^2 \Big ] \Phi^\dagger \Phi \, .
\label{lag.ext}
\end{align}
$\Phi$ denotes the SM Higgs doublet.
 The model describes a SU(2) spontaneously broken theory 
 of massive gauge fields and one physical scalar at high energy.
 The high-energy scale $\Lambda$
 is set by the v.e.v. of the field $\phi$, denoted by $v$.
 The extra gauge fields and the scalar are singlets under the electroweak $SU(2)_L \times U(1)_Y$ group.
 The coupling to the low-energy SM dynamics happens
 via the interaction in the 
 second line of Equation~(\ref{lag.ext}).

 $\lambda$ is the quartic potential
 of the  high-energy scalar field, while $g_1$ is the portal coupling describing the simplest interaction between the high-energy scalar and the Higgs~field.
 
  The singlet gauge fields are
$A_\mu = A_{a\mu} \frac{\tau_a}{2}$,  $\tau_a$ being
the Pauli matrices, while $\phi$ is represented in matrix form by
$\phi~=~\phi_0 + i \tau_a \phi_a$, $\phi_0 = v + \sigma$.
$v$ is the vacuum expectation value of $\phi_0$. 
$\phi_a$ are the high-energy pseudo-Goldstone~fields.

}

Under an infinitesimal gauge transformation of
parameters $\alpha_a$, the fields transform as follows
($g$ is the coupling constant of the extra SU(2) group):
\begin{align}
    & \delta A_{a\mu} = \partial_\mu \alpha_a + 
    g \epsilon_{abc} A_{b\mu} \alpha_c \, ,  \quad
    \delta \phi_a = \frac{g}{2} \phi_0 \alpha_a +
    \frac{g}{2} \epsilon_{abc} \phi_b \alpha_c \, ,
    \quad 
    \delta \phi_0 = -\frac{g}{2} \alpha_a \phi_a \, .
\end{align}
We also define the field strength in the usual way:
\begin{align}
G_{a\mu\nu} = \partial_\mu A_{a\nu} - \partial_\nu A_{a\mu} + g \epsilon_{abc} A_{b\mu} A_{c\nu} \, .
\end{align}
The covariant derivative $D_\mu \phi$ is defined by
\begin{align}
    D_\mu \phi = \partial_\mu \phi 
    -i g A_{a\mu} \frac{\tau_a}{2} \phi \, .
\end{align}
Of course, in~a BSM approach, other operators can be considered, yet the essence of our analysis is unaffected by the particular choice of such operators, so we will limit ourselves to the simplest case
of Equation~(\ref{lag.ext}).

This theory is an example of the so-called Higgs-portal models, see, e.g.,~refs.~\cite{Englert:2011yb,Batell:2011pz,Djouadi:2011aa,Lopez-Honorez:2012tov,Jaeckel:2012yz,Englert:2013gz,LHCHiggsCrossSectionWorkingGroup:2013rie,Buchalla:2013rka,Lopez-Val:2013yba,Beniwal:2015sdl,Robens:2015gla}.
The extended sector affects
the low-energy physics via loop effects, so one might hope
to extract some signals of BSM physics via non-decoupling effects
(i.e., contributions that survive in the large mass limit of the
extra BSM particles).

If only power-counting renormalizable interactions are allowed,
a detailed analysis of these models can be consistently studied~\cite{Robens:2015gla,Arco:2023sac,Dittmaier:2021fls}.

Yet there is no fundamental reason why only dimension four
operators should enter into BSM~physics.

In the present study, we will show that, if some suitable
dimension six and eight operators are introduced, non-decoupling effects vanish once the resummed amplitudes are~considered.

\section{Gauge-Invariant Variables}\label{sec.giv}

In a recent series of papers~\cite{Quadri:2016wwl,Binosi:2017ubk,Binosi:2019olm,Binosi:2019nwz,Binosi:2020unh,Binosi:2022ycu},
the construction of a dynamical gauge-invariant field for the scalar mode was presented.
{For the sake of completeness, we report here a detailed
discussion of the main~results.

The key idea is to introduce
quantum fields that 
are gauge-invariant and are in one-to-one correspondence with the original fields of the~theory. 

Let us consider the field $\phi$.
Its gauge-invariant counter-part is
the composite operator
\begin{align}
    \frac{1}{4v} {\rm Tr} (\phi^\dagger \phi) - \frac{v}{2}  = \sigma + \dots
    \label{scalar.comp.op}
\end{align}
where dots stand for higher dimensional terms in the fields.
The combination in the l.h.s. of
the above equation is gauge-invariant and the normalization is chosen in such a way that, in the linearized approximation, it reduces to the real scalar $\sigma$.

One could try to study a model where
the path-integral is carried out over the original fields $\sigma, \phi_a$, while
the composite operator in Equation~(\ref{scalar.comp.op}) is defined by coupling it to an external source $\beta_{\phi^\dagger \phi}$.
This approach has been widely studied in refs.~\cite{Dudal:2019pyg,Dudal:2021dec,Dudal:2023jsu}.

On the other hand, one could try to construct a model where the 
gauge-invariant field becomes propagating, i.e.,~the path-integral is carried out over those fields. 
For that purpose, one makes use of the Lagrange multiplier technique
    as presented in Refs.~\cite{Quadri:2016wwl,Binosi:2017ubk,Binosi:2019olm,Binosi:2019nwz,Binosi:2020unh,Binosi:2022ycu}.

Therefore, we introduce the gauge-invariant field $h$ together with the Lagrange multiplier field $X$.
By being on-shell
 with $X$, $h$ must reduce to the gauge-invariant combination in the l.h.s. of Equation~(\ref{scalar.comp.op}), i.e.,
}
\begin{align}
    h \sim  \frac{1}{4v} {\rm Tr} (\phi^\dagger \phi) - \frac{v}{2}  = \sigma + \dots
    \label{scalar.equiv}
\end{align}
where the dots stand for higher dimensional terms in the fields
and $\sim$ denotes on-shell~equivalence.

{
In the relevant scalar sector,
the path integral is originally over $\sigma, \phi_a$; then, 
it must be carried out over 
$\sigma, \phi_a$ and additionally  $h, X$.

This procedure is best implemented in a standard way by the Lagrange multiplier technique in the BRST formalism.
For that purpose,}
we introduce the set  $\bar c, c$ of antighost and ghost fields~\cite{Binosi:2022ycu} associated with the above field redefinition.
The  constraint BRST differential $\s$ reads
\begin{align}
    \s \bar c = h -  \frac{1}{4v} {\rm Tr} (\phi^\dagger \phi) + \frac{v}{2}  \, , \quad
    \s X = c \, , \quad \s c = 0 \, , \quad 
    \s h = 0 \, .
\end{align}
Notice that $\s$ is nilpotent due to the gauge invariance of the right hand side (r.h.s.) of Equation~(\ref{scalar.equiv}). It anticommutes
with the ordinary BRST differential $s$ associated with the gauge group.
Under $s$, all fields
$X,h,\bar c, c$ are invariant,
since they are singlet under
both the electroweak 
$SU(2)_L \times U(1)_Y$ group
and the high-energy BSM $SU(2)$ group.
We denote by $s' = s + \s$ the full BRST differential of the~model.

Then, we add to the action of the 
model in the conventional formalism the following BRST-exact 
term ($m^2= 4 \lambda v^2$ is the mass of the $\sigma$-field
in Equation
~(\ref{lag.ext})):
\begin{align}
S_{\tiny{\mbox{aux,scalar}}} & =
\s \int d^4 x \, \Big [ \bar c (\square + m^2) X \Big ] 
=  s' \int d^4 x \, \Big [ \bar c (\square + m^2) X \Big ] 
\nonumber \\
& =
\int d^4 x \, \Big \{
X (\square + m^2) \Big [ h -  \frac{1}{4v} {\rm Tr} (\phi^\dagger \phi) + \frac{v}{2}  \Big ] 
- \bar c (\square + m^2) c 
\Big \} \, .
\label{aux.scalar}
\end{align}
{ 
The (local) physical observables
of the theory are identified
by the cohomology of the full BRST differential $s'$ in the sector with zero ghost number, i.e.,
two operators ${\cal O}, {\cal O}'$ are physically equivalent if they differ by 
a BRST-exact term $s' {\cal R}$,
${\cal O}' = {\cal O} + s' {\cal R}$~\cite{Barnich:2000zw}.

The physical content of the theory is therefore not affected by 
the introduction of $S_{\tiny{\mbox{aux,vect}}}$
in Equation~(\ref{X.term}),
since the latter is a BRST-exact term with ghost number~zero.

Integrating out the extra degrees of freedom $h, \bar c, c$ in the path-integral provides a useful insight into the mechanism at work.
One finds
\begin{align}
\int \, {\cal D} \bar c
{\cal D} c
{\cal D} X
\exp \Big ( i S_{\tiny{\mbox{aux,scalar}}} & \Big ) = 
\int \, {\cal D} \bar c
{\cal D} c
\exp \Big ( - i \int d^4x \, \bar c (\square + m^2) c  \Big )
\nonumber \\
&  \times 
\int {\cal D} X
\exp \Big ( i \int d^4x \, \, X (\square + m^2)
    \Big [ h - \frac{1}{4v} {\rm Tr} (\phi^\dagger \phi) + \frac{v}{2}  \Big ] \Big )
    \nonumber \\
& \sim 
\int \, {\cal D} \bar c
{\cal D} c 
\exp \Big ( - i \int d^4 \, \bar c  (\square + m^2) c \Big ) \nonumber \\
& \qquad \times
\delta \Big [ (\square + m^2)
    \Big [ h - \frac{1}{4v} {\rm Tr} (\phi^\dagger \phi) + \frac{v}{2} \Big ]
\nonumber \\
& \sim \det (\square + m^2) )
\frac{1}{| \det (\square + m^2) | }
\delta \Big [   h - \frac{1}{4v} {\rm Tr} (\phi^\dagger \phi) + \frac{v}{2} \Big ]
\nonumber \\
& \sim \delta \Big [   h - \frac{1}{4v} {\rm Tr} (\phi^\dagger \phi) + \frac{v}{2} \Big ] \, ,
\label{det}
\end{align}
i.e., the determinant arising from the $\delta$-function
is exactly compensated (modulo inessential multiplicative factors) by the integration over the Grassmann variables $\bar c, c$.

}

{
Another way to approach the problem is to go on shell with $X$ in Equation~(\ref{aux.scalar}).}
One obtains a Klein--Gordon equation:
\begin{align}
(\square + m^2) \Big [ h -  \frac{1}{4v} {\rm Tr} (\phi^\dagger \phi) + \frac{v}{2} \Big ] \Rightarrow h = \frac{1}{4v}  
{\rm Tr} (\phi^\dagger \phi) - \frac{v}{2} + \eta \, ,
\label{scalar.sol}
\end{align}
with $\eta$ being a free scalar field with squared mass m$^2$ whose correlators can be proven to vanish
in perturbation theory~\cite{Binosi:2019olm}. 
{ Formally, this
follows from the cancellation
of the determinants in Equation~(\ref{det}), giving back
the constraint in the last
of Equations~(\ref{det}).
A detailed rigorous analysis
of the on-shell reduction is presented in ref.~\cite{Binosi:2019olm}.
}
For this reason,
the $\eta$ mode can be safely set to zero in the following~discussion.

{ So far, the 
introduction of the extra fields
$h,X,\bar c, c$ has not changed the physical content of the model.
One can prove that the physical observables are exactly the same and the $S$-matrix elements are those of the original theory, as follows from the previously reported general cohomological argument that the classical action has been modified by a BRST-exact term.
Diagrammatically, this can be understood as follows.
One can diagonalize the quadratic part in the $\sigma, X$-sector by the linear field redefinition
\begin{align}
    \sigma = \sigma' + X_1 + h \, .
\end{align}
The propagators of $\sigma'$ and $X_1$ have an overall opposite sign (see Equation~(\ref{diag.mass})).
Diagrams involving 
$\sigma'$ and $X_1$
cancel out against each other. The~remaining diagrams give rise to different off-shell amplitudes than in the conventional formalism, yet once one changes to
 on-shell, one recovers the same physical observables as the original theory, as~guaranteed by the BRST invariance of the theory and by the fact that $S_{aux,scalar}$ is BRST-exact.
}

{
Now, the biggest advantage of the extra field formalism becomes~apparent.

We remark that quadratic terms in the gauge-invariant field $h$ will only affect the propagator of $h$ (remember that $h$ is a propagating field, over~which the path-integral is carried out) and thus one can derive a scaling equation 
for the one-particle (1-PI) irreducible amplitudes
by introducing a suitable differential operator
whose eigenvector is the $h$-propagator itself.
The differential operator characterizes the
$h$-propagator as a homogeneous Euler function of weight $-1$, while 1-PI amplitudes with $\lS$ $h$-propagators
will be Euler functions of weight $-\lS$.

Adding such quadratic terms amounts, of course, to changing the physical content of the theory.
The corresponding operators in the conventional formalism are obtained by being on-shell with the extra fields and these give rise to complicated dimension six operators that affect both the quadratic terms and the interaction vertices.
For those operators, resummation in their BSM couplings becomes a hard (if not impossible) task,
while, if one uses gauge-invariant fields, resummation in the BSM couplings under discussion is a simple consequence of the scaling differential equations, as~will be shown in Section~\ref{sec.limit}.
}

To be more specific, one can add the quadratic mass and kinetic terms to the
classical action
\begin{align}
\int d^4 x\Big ( 
-\frac{M^2-m^2}{2} h^2 - \frac{z}{2} h \square h \Big ) \, .
\label{scalar.mass.kin}
\end{align}
These are physical gauge-invariant operators and 
they modify the physical content of the~theory.

By substituting back the solution for $h$ in Equation~(\ref{scalar.sol}) 
at $\eta=0$  into the classical vertex functional
Equation~(\ref{cl.act}), we obtain their counter-parts in the conventional formalism:
\begin{align}
\int d^4 x\Big [ 
-\frac{M^2}{32 v^2} \Big ( {\rm Tr} (\phi^\dagger \phi) - 2v \Big )^2 
- \frac{z}{32 v^2} {\rm Tr} (\phi^\dagger \phi) \square {\rm Tr} (\phi^\dagger \phi) \Big ] \, .
\label{scalar.mass.kin.trad}
\end{align}
We remark that the $m^2$-dependent term cancels out against the corresponding contribution in the classical action in Equation~(\ref{cl.act}), i.e.,~the only physical parameters are $M$ and $z$.
The cancellations involving $m^2$ have been discussed 
in ref.~\cite{Binosi:2019olm}.

Notice that a dimension six operator has appeared via the kinetic term in $h$. This will play a crucial role in the construction of the decoupling~limit.

As anticipated, the ordinary formalism 
operators in Equations~(\ref{scalar.mass.kin.trad}) yield a complicated set of interactions that is hard to treat beyond the small coupling regime.
On the other hand, the~scaling differential equation for the operators in Equation~(\ref{scalar.mass.kin.trad}) will give us, for free, the resummation in $z$ (and important phenomenological consequences for the SM effective field theory program, as~we will explain in the next sections).

\subsection{Gauge Field}

We now move to the construction of a dynamical gauge-invariant variable for the massive gauge field $A_{a\mu}$.
For the sake of simplicity, we will consider the Landau gauge.
The complete analysis in an arbitrary $R_\xi$-gauge will be presented~elsewhere.

In order to set the stage, we first need to fix the gauge
{\em à la} BRST,
so we add to the Lagrangian in Equation~(\ref{lag.ext}) the following gauge-fixing term:
\begin{align}
    S_{\tiny{\mbox{g.f. + ghost}}} = 
    \int d^4 x\Big [ - b_a \partial A_a + \bar c_a \partial^\mu D_\mu c_a \Big ] \, . 
\end{align}
The covariant derivative acts on the ghost fields $c_a$ as
\begin{align}
    D_\mu c_a = \partial_\mu c_a + g \epsilon_{abc} A_{b\mu} c_c \, .
\end{align}
The relevant gauge-invariant counter-part of the gauge field $A_\mu$ is
\begin{align}
a_\mu & \sim  \frac{i}{g v^2} \Big [ 2 \phi^\dagger  D_\mu \phi - \partial_\mu ( \phi^\dagger \phi ) \Big ] \nonumber \\
& = \Big ( A_{a\mu} - \frac{2}{gv} \partial_\mu \phi_a \Big ) \tau_a + \dots
\label{g.inv.a}
\end{align}
where the dots stand for terms of higher dimension in the~fields.

The procedure to enforce the on-shell constraint in Equation~(\ref{g.inv.a}) follows the same lines as in the scalar case.
The additional anti-ghost $\bar c_\mu$ is now a vector field,
$\bar c_\mu = \bar c_{a\mu} \frac{\tau_a}{2}$,
transforming under the constraint BRST differential $\s$ as
\begin{align}
    \s \bar c_\mu = a_\mu - \frac{i}{gv^2} \Big [ 2 \phi^\dagger  D_\mu \phi - \partial_\mu ( \phi^\dagger \phi ) \Big ] \, .
\end{align}
The constraint ghost and Lagrange multiplier are, respectively,
$c_\mu = c_{a\mu} \frac{\tau_a}{2}$ and
$X_\mu = X_{a\mu} \frac{\tau_a}{2}$.
They form a BRST doublet~\cite{Gomis:1994he,Barnich:2000zw,Quadri:2002nh} under $\s$:
\begin{align}
    \s X_\mu = c_\mu \, , \qquad
    \s c_\mu = 0 \, .
\end{align}
The nilpotency of $\s$ again follows
from the gauge invariance of the r.h.s. of Equation~(\ref{g.inv.a}). 

The additional terms to be added to the action are
\begin{align}
    S_{\tiny{\mbox{aux,vect}}} & = \int d^4 x\, \s {\rm Tr} \Big (
    \bar c_\mu \Sigma^{\mu\nu} X_\nu
    \Big ) \nonumber \\
    & = \int d^4 x\, {\rm Tr} \Big \{ - \bar c_\mu \Sigma^{\mu\nu} c_\nu +
    X_\mu \Sigma^{\mu\nu} 
    \Big [ a_\nu - \frac{i}{gv^2} \Big ( 2 \phi^\dagger  D_\nu \phi - \partial_\nu ( \phi^\dagger \phi ) \Big ) \Big ] 
    \Big \} \, ,
    \label{X.term}
\end{align}
where the symmetric tensor $\Sigma^{\mu\nu}$
denotes the two-point 1-PI amplitude of the gauge field $A_\mu$ in the Landau gauge and is given by
\begin{align}
  \Sigma^{\mu\nu} = 
(\square g^{\mu\nu} - \partial^\mu \partial^\nu) + M_A^2 g^{\mu\nu} \, ,
\end{align}
with $M_A = g v/2$ being the mass of the vector~field.

In the Landau gauge, the gauge field propagator is transverse and the pseudo-Goldstone field is massless. The physical unitarity in this gauge has been studied in detail in ref.~\cite{Ferrari:2004pd}.

The quadratic part in the relevant sector reads
\begin{align}
    \int d^4 x\Big [ 
    \frac{1}{2} A_{a\mu} ( \square g^{\mu\nu} - \partial^\mu \partial^\nu ) A_{a\nu} + \frac{M_A^2}{2} \Big ( A_{a\mu} - \frac{1}{M_A} \partial_\mu \phi_a \Big )^2 - b_a \partial A_a \Big ] \, .
    \label{quad.landau}
\end{align}

The propagators can be obtained by diagonalizing the two-point 1-PI amplitudes in the sector spanned by $A_{a\mu}, a_{a\mu}, X_{a\mu}, b_a, \phi_a$.
The derivation is presented in Appendix~\ref{app:landau}.

We notice that the
mass eigenstate $a'_{a\mu}$
in Equation~(\ref{loc.field.redef.2})
is also BRST-invariant,
since according to Equation~(\ref{loc.field.redef.0}),  it is given by
\begin{align}
a'_{a\mu} = a_{a\mu} - \frac{1}{M_A^2} \partial_\mu b_a \, ,
\end{align}
i.e., a linear combination of gauge-invariant variables.
Hence, one can freely add an independent mass term
\begin{align}
\int d^4 x
\frac{\Mp2- M_A^2}{2} {a'_{a\mu}}^2 
\label{new.mass}
\end{align}
as well as a transverse combination
\begin{align}
    \int d^4 x\frac{z'}{2} a'_{a\mu} (\square g^{\mu\nu} - \partial^\mu \partial^\nu) a'_{a\nu}
    \label{new.kin}
\end{align}
while preserving gauge-invariance. $M'$ and $z'$ are additional BSM couplings, as~well as $M$ and $z$ in Equation~(\ref{scalar.mass.kin}).

Other choices involving the longitudinal parts are also possible
(e.g., $(\partial a'_a)^2$), yet when these operators are switched on and one moves to on-shell, quadratic higher derivative terms in the pseudo-Goldstone fields arise
and, consequently, negative norm states are theoretically
 introduced~\cite{Langlois:2015skt,Ostrogradsky:1850fid}.
For this reason, we limit to the contributions in
Equations~(\ref{new.mass}) and (\ref{new.kin}).

The $a'_\mu$-propagator is correspondingly modified as 
\begin{align}
    \Delta_{a'_{a\mu} a'_{b\nu}} = \frac{i \delta_{ab}}{-(1+z')p^2 + \Mp2} T_{\mu\nu} + \frac{i \delta_{ab}}{\Mp2}L_{\mu\nu} \, . 
\end{align}
while all other propagators are~unaffected.

Even at $z'=0$, the shift in the mass term induces a violation of power-counting renormalizability, since now the $A_\mu$-propagator develops a constant
longitudinal part
\begin{align}
    \Delta_{A_{a\mu} A_{b\nu}} = \frac{i \delta_{ab}}{-p^2 +  \Mp2} T_{\mu\nu} +
    i \delta_{ab} \frac{  \Mp2 - M_A^2}{M_A^2 \Mp2} L_{\mu\nu}
\end{align}
unless $M' = M_A$. 

The violation of power-counting renormalizability by the $a'_\mu$-mass term can be understood by noticing that there are two contributions to the mass term 
%

\begin{align}
    \int d^4 x\frac{\Mp2 - M_A^2}{2} {a'_{a\mu}}^2 & = 
    \int d^4 x\frac{\Mp2  - M_A^2}{2} \Big ( a_{a\mu} - \frac{1}{M_A^2} \partial_\mu b_a 
    \Big )^2 \nonumber \\
    & =  \int d^4 x\Big ( \frac{\Mp2 - M_A^2}{2} a_{a\mu}^2 - 
    \frac{\Mp2 - M_A^2}{M_A^2}a_{a\mu} \partial^\mu b_a +
    \frac{\Mp2 - M_A^2}{2 M_A^4} \partial^\mu b_a \partial_\mu b_a \Big ) \, .
\end{align}

As a consequence of the gauge invariance of $a_\mu$, 
the last two terms in the above equation can be removed by adding
the BRST-exact term
\begin{align}
    \frac{\Mp2 - M_A^2}{M_A^2} \int d^4 x\s \Big [
    \partial_\mu \bar c_a 
    \Big ( a_{a\mu} - \frac{1}{2 M_A^2}
    \partial_\mu b_a \Big )
    \Big ] \, . 
\end{align}
and they are thus unphysical.
The first term is the on-shell equivalent of the dim.6 operator
\begin{align}
\int d^4 x\, & \frac{\Mp2 - M_A^2}{2} {a_{a\mu}}^2 =
\int d^4 x(\Mp2 - M_A^2) ~{\rm Tr } ~a_\mu^2\sim
\nonumber \\
& \frac{\Mp2 - M_A^2}{4 v^2 M_A^2} \int d^4 x\, {\rm Tr} \, \Big \{ 
\phi^\dagger \phi \Big [ 4 D^\mu \phi \,  (D_\mu \phi)^\dagger +
2 \partial^\mu ( \phi^\dagger D_\mu \phi + (D_\mu \phi)^\dagger \phi ) - \square \phi^\dagger \phi
\Big ] \Big \} \, .
\label{non.ren.term}
\end{align}
The classical action is thus modified by a non-renormalizable interaction.
The relevant term giving a mass contribution to the gauge field
is the first one in the r.h.s.
of Equation~(\ref{non.ren.term}), belonging to the family of operators
\begin{align}
    C_n \equiv \int d^4 x{\rm Tr} \,  [ (\phi^\dagger \phi)^n (D^\mu \phi)^\dagger D_\mu \phi ] \, .
\end{align}
All of them contribute to the
gauge field mass term. As~is very well known, only $C_0$ leads to a
power-counting renormalizable~theory.

By the same argument, the additional kinetic term corresponds to a dimension eight operator
with four derivatives
\begin{align}
    \int d^4 x\, \frac{z'}{2} a'_{a\mu} (\square g^{\mu\nu} - \partial^\mu \partial^\nu) a'_{a\nu} \sim
    - \frac{4 z'}{g^2 v^4} 
    \int d^4 x{\rm Tr} \, \Big [ \phi^\dagger D_\mu \phi 
    (\square g^{\mu\nu} - \partial^\mu \partial^\nu) (\phi^\dagger D_\nu \phi ) \Big ] \, .
    \label{vect.kin.trad}
\end{align}
It contributes both to the quadratic part and to the interaction terms, as~also happens
for the terms in Equation~(\ref{non.ren.term}).

In the standard formalism, it is difficult to compute the radiative corrections induced by those operators beyond the small coupling expansion and it is very hard to guess the form of the~resummation.

On the other hand, by~using the dynamical gauge-invariant fields, the additional operators
are rewritten in a form that only contributes to the quadratic~part.

This paves the way for the derivation of some
novel differential equations, allowing for the determination of the functional dependence of
the amplitudes on the new parameters in an exact way. This will be discussed in the next~section.

\section{The Decoupling Limit}\label{sec.limit}

The parameters $z, M^2$ and $ z', \Mp2$ only enter in the  propagators 
$\Delta_{hh}$ and $\Delta_{a'_{a\mu} a'_{b\mu}}$, respectively, and never in the interaction vertices.
Moreover, the propagator $\Delta_{hh}$ is an eigenvector of
eigenvalue $-1$ of the differential operator
\begin{align}
    \dEulS \equiv (1+z) \frac{\partial}{\partial z} + M^2 \frac{\partial}{\partial M^2} \, ,
\end{align}
while 
the propagator $\Delta_{a'_{a\mu} a'_{b\mu}}$ is an eigenvector of eigenvalue $-1$ of the differential operator
\begin{align}
    \dEulV \equiv (1+z') \frac{\partial}{\partial z'} + \Mp2 \frac{\partial}{\partial \Mp2} \, ,
\end{align}
i.e.,
\begin{align}
    \dEulS \Delta_{hh} = - \Delta_{hh} \, , \qquad
    \dEulV \Delta_{a'_{a\mu} a'_{b\mu}} = - \Delta_{a'_{a\mu} a'_{b\mu}} \, .
\end{align}
Let us now consider an $n$-th loop 1-Pi amplitude $\G^{(n)}_{\varphi_1 \dots \varphi_r}$ with $r$ $\varphi_i$
external legs, $\varphi_i = \varphi(p_i)$ denoting
a generic field or external source of the theory with incoming momentum $p_i$.

$\G^{(n)}_{\varphi_1 \dots \varphi_r}$ can be decomposed as the sum
of all diagrams with (amputated) external legs $\varphi_1 \dots \varphi_r$ with  zero, one, two, $\dots$ $\lS$ internal
$h$-propagators
and
zero, one, two, $\dots$ $\lV$ internal
$a'$-propagators:
\begin{align}
    \G^{(n)}_{\varphi_1 \dots \varphi_r} = \sum_{\lS,\lV \geq 0}
    \G^{(n; \lS, \lV)}_{\varphi_1 \dots \varphi_r}
\end{align}

Then, each $\G^{(n; \lS, \lV)}_{\varphi_1 \dots \varphi_r}$ is clearly an eigenvector of $\dEulS$ of eigenvalue $-\lS$  and of
$\dEulV$ of eigenvalue
$-\lV$,
namely
\begin{align}
    \dEulS \G^{(n;  \lS, \lV)}_{\varphi_1 \dots \varphi_r} = 
    - \lS \G^{(n; \lS, \lV)}_{\varphi_1 \dots \varphi_r}\, , 
    \qquad
    \dEulV \G^{(n;  \lS, \lV)}_{\varphi_1 \dots \varphi_r} = 
    - \lV \G^{(n; \lS, \lV)}_{\varphi_1 \dots \varphi_r}\, .  
\end{align}
According to Euler's theorem, the most general solution to the above differential equations
is a homogeneous function in the variables $M^2/(1+z)$ and $\Mp2/(1+z')$, i.e.,~a function of the form
\begin{align}
\G^{(n; \lS, \lV)}_{\varphi_1 \dots \varphi_r} (z, M^2, z', \Mp2) = 
 \frac{1}{(1+z)^{\lS}} \frac{1}{(1+z')^{\lV}}\G^{(n; \lS, \lV)}_{\varphi_1 \dots \varphi_r} \Big ( 0, \frac{M^2}{1+z}, 0, \frac{\Mp2}{1+z'} \Big ) \, .
 \label{scaling}
\end{align}
Notice that this result holds true for all orders in the loop~expansion.

This is preserved by renormalization, provided that the finite normalization conditions are chosen in such a way to fulfil Equation~(\ref{scaling})~\cite{Binosi:2022ycu}.

Equation~(\ref{scaling}) predicts the structure of the fully resummed amplitudes, i.e.,~it contains the exact dependence on the parameters $z,z',M^2, \Mp2$.

In particular, one can consider trajectories in the couplings space where
the ratios $M^2/(1+z)$ and $\Mp2 /(1+z')$ are kept fixed while letting $z,z'$ tend towards~infinity.

According to Equation~(\ref{scaling}), one sees that, in such a limit, only the contribution
$\G^{(n;0,0)}$ will survive.
This is equivalent to saying that all contributions generated by diagrams, where at least one internal line is a $\Delta_{hh}$ or a$\Delta_{a'a'}$-propagator, vanish, i.e.,~all non-decoupling effects are washed~out.

This is quite a surprising result. In~fact, it implies that the high-energy
dynamics are totally decoupled from the infrared regime. In~some sense,
physical particles of the high-energy theory act as classical background sources, influencing the low-energy physics only by tree-level~contributions.

{It is worth commenting
on the potential phenomenological impact of such  decoupling limits on 
BSM~fits.

The EFT description of the low-energy effects, arising from integrating out the gauge and scalar fields of the extended Lagrangian (\ref{lag.ext}), can be arranged as a set of operators of increasing dimension in the high-energy scale $\Lambda$.

In the present case, the typical energy scale $\Lambda$ is set by $v$, i.e.,~the scale controlling the spontaneous symmetry breaking in the high-energy~sector.

The EFT description will give rise to an effective low-energy Lagrangian of the form
\begin{align}
{\cal L}_{eff} = {\cal L}_{SM} + \frac{1}{\Lambda} {\cal L}_5 + \frac{1}{\Lambda^2} {\cal L}_6 +
\frac{1}{\Lambda^3} {\cal L}_7 + \frac{1}{\Lambda^4} {\cal L}_8 + \dots, \qquad
{\cal L}_d = \sum_i c_i^{(d)} {\cal cal O}^{(d)}_i \, .
\end{align}
${\cal O}^{(d)}_i$ are local $\mbox{SU(3)}_c \times \mbox{SU(2)}_L \times \mbox{U(1)} _Y$-invariant operators of canonical dimension $d$, involving only the light SM~fields.

${\cal L}_{SM}$ is the SM Lagrangian, which is known to be a very good description of electroweak physics, with~the exception of the small neutrino masses that are taken into account by ${\cal L}_5$.

The higher order terms ${\cal L}_{d \geq 5}$ yield contributions to physical processes that are suppressed by factors $(E/\Lambda)^{d-4}$, where $E$ is the relevant energy scale of the process under investigation.
The virtual effects due to the propagation of the heavy degrees of freedom inside loops are captured by
the Wilson coefficients $c_i^{(d)}$.
The latter can be computed by matching with the UV complete high-energy~theory.

The effects described by ${\cal L}_{d \geq 5}$ must be small, due to the success of ${\cal L}_{SM}$ in describing experimental~data.

This hints at the fact that either the scale of new physics  
$\Lambda$ is very large or the structure of ${\cal L}_{d \geq 5}$ is
particularly elaborated, or~perhaps a combination of both~\cite{deBlas:2019rxi}.

In the standard treatment of SM Effective Field Theories, one takes the assumption that $\Lambda$ is large and carries out a perturbative treatment of amplitudes in the small coupling regimes with respect to the BSM~parameters.

The mechanism described in the present study offers a precise new way to
ensure that the effects of ${\cal L}_{d \geq 5}$ are small based on resummation; Equation~(\ref{scaling}) implies
that the full amplitudes (beyond the small coupling approximation in $z,M^2,z',M^{'2}$) are suppressed
in the {\em strong} coupling regime for $z,z'$.

This is a non-trivial result that points towards the necessity of going beyond the small coupling expansion in the SM effective field theory~program.

In view of the fact that
HL-LHC will allow for studying
anomalous couplings with a precision
of a few percent, and that
future colliders can significantly improve that precision~\cite{deBlas:2019rxi,deBlas:2022ofj},
a comparison with the experimental data 
of the SM effective field theory fits must take into account the resummation effects.}

\section{Conclusions}\label{sec.conclusions}

In the present study, we have shown how to construct dynamical
gauge-invariant variables for the gauge fields, by~extending
the procedure already obtained for scalars.
The method works whenever spontaneous symmetry breaking~occurs.

Dynamical gauge-invariant variables are quantum fields
over which the path-integral is carried out, at~variance with
the approach based on composite gauge-invariant operators~\cite{Dudal:2023jsu,Dudal:2021dec,Dudal:2019pyg}.

One of the main advantages of the gauge-invariant dynamical fields
is that they allow for representing certain operators, involving complicated interactions in the standard formalism, by~purely
quadratic contributions to the classical~action.

This, in turn, allows one to derive powerful differential equations
controlling the dependence of the 1-PI amplitudes on their~coefficients.

For the special choice of dimension six and dimension eight
two derivative operators for the scalar and the vector fields,
given in Equations~(\ref{scalar.mass.kin}), (\ref{new.mass}) and (\ref{new.kin}), (respectively, and
Equations~(\ref{scalar.mass.kin.trad}), (\ref{non.ren.term}) and (\ref{vect.kin.trad}) in the ordinary formalism), this implies that fully resummed amplitudes
have a fairly simple dependence on such BSM couplings (see
Equation~(\ref{scaling})).

Then, it is easy to identify trajectories in the coupling space where non-decoupling effects from the propagation of high-energy particles are washed out, while keeping the (tree-level) pole masses of such particles at fixed values (constant ratios $M^2/(1+z)$ and
$\Mp2/(1+z')$).

The argument is very general and does not depend on the particular interactions in the high-energy theory. It applies whenever the spectrum only contains ordinary particles (no higher derivatives terms in the quadratic part) and when operators
exist in the high-energy theory, such as those given in
Equations~(\ref{scalar.mass.kin.trad}), (\ref{non.ren.term}) and (\ref{vect.kin.trad}) in the standard~formalism.

\vspace{6pt} 





\section*{The author acknowledges the warm hospitality of the Mainz Institute for Theoretical Physics (MITP) during the MITP Scientific Program ``EFT Foundations and Tools'' in August 2023, where part of this work was carried out.}

\appendix
\section{The Model and Its Symmetries}\label{app.model}

The complete classical vertex functional reads
%

\begin{align}
    \G^{(0)} =  \G^{(0)}_{\mbox{SM}} + \int d^4 x\Big \{ 
    & -\frac{1}{4} G_{a\mu\nu}^2 +
\frac{1}{4} 
{\rm Tr}
(D^\mu \phi)^\dagger D_\mu \phi -
\frac{\lambda}{2} \Big [ \frac{1}{2} {\rm Tr} (\phi^\dagger \phi) - v^2 \Big ]^2 \nonumber \\
& + g_1 {\rm Tr} \Big [ (\phi^\dagger \phi) - v^2 \Big ]^2 \Phi^\dagger \Phi 
\nonumber \\
&
-\frac{M^2-m^2}{2} h^2 - \frac{z}{2} h \square h 
+ \frac{M_a^2 - M_A^2}{2} {a'_\mu}^2 + \frac{z_a}{2} a'_\mu (\square g^{\mu\nu} - \partial^\mu \partial^\nu) a'_\nu \nonumber \\
& 
- b_a \partial A_a + \bar c_a \partial^\mu D_\mu c_a \nonumber \\
& + X (\square + m^2) \Big [ h -  \frac{1}{4v} {\rm Tr} (\phi^\dagger \phi) + \frac{v}{2}  \Big ] 
- \bar c (\square + m^2) c \nonumber \\
& +
{\rm Tr} \Big \{
    X_\mu \Sigma^{\mu\nu} 
    \Big [ a_\nu - \frac{i}{gv^2} \Big ( 2 \phi^\dagger  D_\nu \phi - \partial_\nu ( \phi^\dagger \phi ) \Big ) \Big ]  - \bar c_\mu \Sigma^{\mu\nu} c_\nu 
    \Big \} \nonumber \\
    & + \bar c^* \Big ( h -  \frac{1}{4v} {\rm Tr} (\phi^\dagger \phi) + \frac{v}{2} \Big ) 
    + {\rm Tr} \Big \{ \bar c^{*\mu } \Big [  a_\mu - \frac{i}{gv^2} \Big ( 2 \phi^\dagger  D_\mu \phi - \partial_\mu ( \phi^\dagger \phi ) \Big ) 
    \Big ] \nonumber \\
    & + A^{* \mu}_a D_\mu c_a -  \frac{g}{2} \phi^*_0 c_a \phi_a  
    + \phi_a^* \Big ( \frac{g}{2} \phi_0 c_a + \frac{g}{2} \epsilon_{abc} \phi_b \omega_c \Big )
\Big \} \, .
\label{cl.act}
\end{align}

In the above equation, $\G^{(0)}_{\mbox{SM}}$ denotes the classical vertex functional of the SM, including the SM classical action, the~gauge-fixing and ghost terms, as well as the external sources (the so-called antifields~\cite{Gomis:1994he}) required to define, at the quantum level, the BRST
transformations of the fields generated by the electroweak gauge~group.

The additional higher dimensional operators written in terms of
gauge-invariant dynamical fields are reported in the third line
of Equation~(\ref{cl.act}).

$a'_\mu$ is defined by
$$a'_{a\mu} = a_{a\mu} - \frac{1}{M_A^2} \partial_\mu b_a \, . $$
The fourth and fifth lines contain the Lagrange multipliers together with the constraint ghost and antighost fields (that remain free).
The differential operator $\Sigma^{\mu\nu}$ is
\begin{align}
  \Sigma^{\mu\nu} = 
(\square g^{\mu\nu} - \partial^\mu \partial^\nu) + M_A^2 g^{\mu\nu} \, .
\end{align}

Finally, in the last two lines, the antifields for the constraint BRST
transformations $\s \bar c, \s \bar c_\mu$, as well as for the SU(2) gauge group BSM extension, are introduced with~the
convention
$$\bar c^{*\mu} = \bar c^{* \mu}_a \frac{\tau_a}{2} \, .$$

The classical action in Equation~(\ref{cl.act}) obeys several functional
identities, in~addition to the usual ones (Slavnov--Taylor identities,
$b$-equation, ghost equation) valid for the SM part $\G^{(0)}_{\mbox{SM}}$:
\begin{itemize}
    \item The $X$- and $X_a^\mu$-equations:
\begin{align}
        & \frac{\delta \G^{(0)}}{\delta X } = (\square + m^2) \frac{\delta \G^{(0)}} {\delta \bar c^*} \, , \nonumber \\
        & \frac{\delta \G^{(0)}}{\delta X_a^\mu } = \frac{1}{2} \Sigma^{\mu\nu} \frac{\delta \G^{(0)}}{\delta \bar c^{* \nu}_a} \, ;
        \label{X.eqs}
    \end{align}
    \item The $b$-equation and ghost equation for the high-energy SU(2) gauge group:
\begin{align}
    \frac{\delta \G^{(0)}}{\delta b_a} = \partial A_a \, , \qquad
    \frac{\delta \G^{(0)}}{\delta \bar c_a} = \partial^\mu \frac{\delta \G^{(0)}}{\delta A_a^{*\mu}} \, ;
    \end{align}
    \item The $h$ and $a_\mu$-equations:
\begin{align}
    & \frac{\delta \G^{(0)}}{\delta h} = - (M^2 -m^2) h - z \square h + (\square + m^2) X + \bar c^* \, , \nonumber \\
    & \frac{\delta \G^{(0)}}{\delta a_{a\mu}} = (M_a^2 - M_A^2) a'_{a\mu} + 
    z_a (\square g^{\mu\nu} - \partial^\mu\partial^\nu) a'_{a\nu} +
    \frac{1}{2} \Sigma^{\mu\nu} X_{\nu a} + \bar c^*_{a\mu} \, .
    \end{align}
    Notice that the r.h.s. is linear in the quantum fields and, therefore, no
    further external source is required to renormalize these identities.
    \item The high-energy SU(2) Slavnov--Taylor identity:
\begin{align}
        {\cal S}(\G^{(0)}) = \int d^4 x
        \Big [ \frac{\delta \G^{(0)}}{\delta A^*_{a\mu}}
               \frac{\delta \G^{(0)}}{\delta A_{a\mu}}+
               \frac{\delta \G^{(0)}}{\delta \sigma^*}
               \frac{\delta \G^{(0)}}{\delta \sigma}+
               \frac{\delta \G^{(0)}}{\delta \phi^*_{a}}
            \frac{\delta \G^{(0)}}{\delta \phi_a}+
            b_a \frac{\delta \G^((0)}{\delta \bar c_a}  
        \Big ] = 0 \, ;
     \end{align}       
    \item The $b$-equation
\begin{align}
        \frac{\delta \G^{(0)}}{\delta b_a } = - \partial A_a ;
    \end{align}
    \item The ghost equation
\begin{align}
    \frac{\delta \G^{(0)}}{\delta \bar c_a }= \partial^\mu \frac{\delta \G^{(0)}}{\delta A^*_{a\mu}} ;
    \end{align}
    \item The constraint Slavnov--Taylor identities:
\begin{align}
    & {\cal S}_{\s,{\tiny{\mbox{scal}}}}(\G^{(0)}) = \int d^4 x\Big [ 
    c \frac{\delta \G^{(0)}}{\delta X} + \frac{\delta \G^{(0)}}{\delta \bar c^*}\frac{\delta \G^{(0)}}{\delta \bar c}
    \Big ]  = 0 \, , \nonumber \\
    & {\cal S}_{\s,{\tiny{\mbox{vect}}}}(\G^{(0)}) = \int d^4 x\Big [ 
    c_{a\mu} \frac{\delta \G^{(0)}}{\delta X_{a\mu}} + \frac{\delta \G^{(0)}}{\delta \bar c^*_{a \mu}}\frac{\delta \G^{(0)}}{\delta \bar c_{a\mu}}
    \Big ]  = 0 \, ;
    \label{st.constr}
    \end{align}
    \item The ghost equations for the constraint ghosts:
\begin{align}
        \frac{\delta \G^{(0)}}{\delta \bar c} = -(\square + m^2) c \, , \qquad
        \frac{\delta \G^{(0)}}{\delta \bar c_{a\mu}} = - \frac{1}{2} \Sigma^{\mu\nu} c_a \, .
        \label{ghost.eqs}
    \end{align}
    By using Equation~(\ref{ghost.eqs}) in Equation~(\ref{st.constr}), one obtains the $X$-equations (\ref{X.eqs}).
\end{itemize}

\section{Propagators}\label{app:propagators}
\unskip

\subsection{Scalar Fields}\label{app:scalar}

The diagonalization of the quadratic part in the scalar sector spanned by $\sigma, X, h$  is achieved by setting~\cite{Binosi:2022ycu}
\begin{align}
X = X_1 + h \, , \qquad \sigma = \sigma' + X_1 + h \, .
\end{align}
The propagators in the mass eigenstate (diagonal) basis are
\begin{align}
    \Delta_{\sigma'\sigma'} = - \Delta_{X_1 X_1} = \frac{i}{p^2 - m^2} \, , \qquad
    \Delta_{hh} = \frac{i}{(1+z) p^2 - M^2} \, .
    \label{diag.mass}
\end{align}

\subsection{Landau Gauge}\label{app:landau}

One must diagonalize the quadratic part given by 
\begin{align}
     \int d^4 x\Big \{ & 
    \frac{1}{2} A_{a\mu} ( \square g^{\mu\nu} - \partial^\mu \partial^\nu ) A_{a\nu} + \frac{M_A^2}{2} \Big ( A_{a\mu} - \frac{1}{M_A} \partial_\mu \phi_a \Big )^2 - b_a \partial A_a \nonumber \\ 
    & + X_{a\mu} \Big [ (\square g^{\mu\nu} - \partial^\mu \partial^\nu) + M_A^2 g^{\mu\nu}  \Big ] \Big ( a_{a\nu} - A_{a\nu} + \frac{1}{M_A} \partial_\nu \phi_a \Big ) \Big \} \, .
    \label{quad.landau.0}  
\end{align}
One first removes the $\phi-A_\mu$-mixing via the redefinition
\begin{align}
    b_a = b'_a + M_A \phi_a \, ,
    \label{loc.field.redef.0}
\end{align}
followed by the cancellation of the $b'-A_\mu$-mixing by the replacement
\begin{align}
A_{a\mu} = A'_{a\mu} - \frac{1}{M_A^2}\partial_\mu b'_a \, .
\end{align}
A further set of field redefinitions
\begin{align}
    A'_{a\mu} = A''_{a\mu} + X_{a\mu} \, , \qquad X_{a\mu} = X'_{a\mu} + a_{a\mu} 
\end{align}
take care of the $X_\mu-A_\nu$ and $X_\mu-a_\nu$ mixing. 
One is eventually left with
\begin{align}
     \int d^4 x\Big \{ & 
    \frac{1}{2} A''_{a\mu} [ (\square + M_A^2) g^{\mu\nu} - \partial^\mu \partial^\nu ] A''_{a\nu} 
    -
    \frac{1}{2} X'_{a\mu} [ (\square + M_A^2) g^{\mu\nu} - \partial^\mu \partial^\nu ] X'_{a\nu}
    \nonumber \\
    & +
     \frac{1}{2} a_{a\mu} [ (\square + M_A^2) g^{\mu\nu} - \partial^\mu \partial^\nu ] a_{a\nu} 
  - \frac{1}{2 M_A^2} \partial^\mu b'_a\partial_\mu b'_a
    + \frac{1}{2} \partial^\mu \phi_a \partial_\mu \phi_a   \nonumber \\ 
  &
  + ( X'_{a\mu} + a_{a\mu} ) \partial^\mu (b'_a + M_A \phi_a ) \Big \} \, .  
    \label{quad.landau.1}
\end{align}
The mixing terms in the last line of the above equation can be removed by the local field redefinition 
\begin{align}
    X'_{a\mu} = X''_{a\mu} + \frac{1}{M_A^2} \partial_\mu b'_a + \frac{1}{M_A} \partial_\mu \phi_a \, , \qquad
    a_{a\mu} = a'_{a\mu} -\frac{1}{M_A^2} \partial_\mu b'_a - \frac{1}{M_A} \partial_\mu \phi_a \, .
    \label{loc.field.redef.2}
\end{align}
No new $b'-\phi$-mixing is generated. The~diagonal propagators in momentum space (mass eigenstates) are finally given by
\begin{gather}
    \Delta_{A''_{a\mu} A''_{b\nu}} = \Delta_{a'_{a\mu} a'_{b\nu}} = - \Delta_{X''_{a\mu} X''_{b\nu}} = \frac{i \delta_{ab}}{-p^2 + M_A^2} T_{\mu\nu} + \frac{i\delta_{ab}}{M_A^2}L_{\mu\nu} \, 
    \nonumber \\
    \Delta_{b'_a b'_b} = -\frac{i \delta_{ab} M_A^2}{p^2} \, , ~~
    \Delta_{\phi_a\phi_b} = \frac{i\delta_{ab}}{p^2} \, . 
\end{gather}
In the symmetric basis $(A_\mu, \phi, b, X_\mu, a_\mu)$ 
\begin{align}
& b_a = b'_a + M_A \phi_a \, , \qquad A_{a\mu} = A''_{a\mu}+X''_{a\mu} +a'_{a\mu} - \frac{1}{M_A}\partial_\mu b'_a \, , \nonumber \\
& X_{a\mu} = X''_{a\mu} + a'_{a\mu} \, , \qquad
a_{a\mu} = a'_{a\mu} - \frac{1}{M_A^2}\partial_\mu b'_a - \frac{1}{M_A} \partial_\mu \phi_a 
\label{landau.diag}
\end{align}
the non-vanishing propagators
are given by
\begin{gather}
    \Delta_{A_{a\mu} A_{b\nu}} = \frac{i\delta_{ab}}{-p^2 + M_A^2} T_{\mu\nu} \, , \qquad
    \Delta_{A_{a\mu} b_b} = - \frac{\delta_{ab} p_\mu}{p^2}\, , \qquad
    \Delta_{A_{a\mu} a_{b\nu}} = \frac{i \delta_{ab}}{-p^2 + M_A^2} T_{\mu\nu} \, , \nonumber \\
    \Delta_{X_{a\mu} a_{b\nu}} = \frac{i\delta_{ab}}{-p^2 + M_A^2} T_{\mu\nu} + \frac{i \delta_{ab}}{M_A^2} L_{\mu\nu} \, , \qquad
    \Delta_{a_{a\mu} a_{b\nu}} = \frac{i\delta_{ab}}{-p^2 + M_A^2} T_{\mu\nu} + \frac{i\delta_{ab}}{M_A^2} L_{\mu\nu} \, , \nonumber \\
    \Delta_{a_{a\mu} \phi_b} = \frac{\delta_{ab}}{M_A}\frac{p_\mu}{p^2} \, , \qquad
    \Delta_{b_a \phi_b} = \frac{i \delta_{ab} M_A}{p^2} \, , \qquad
    \Delta_{\phi_a\phi_b} = \frac{i \delta_{ab}}{p^2} \, .
    \label{landau.prop.symm}
\end{gather}

\end{document}